\documentclass[runningheads]{llncs}
\usepackage[T1]{fontenc}
\usepackage{graphicx}
\usepackage{multirow}
\usepackage{hyperref}
\usepackage{xcolor}

\begin{document}
\title{A 3D generative model of pathological multi-modal MR images and segmentations}
\titlerunning{A 3D generative of pathological multi-modal MR images}

\author{Virginia Fernandez\inst{1}\orcidID{0000-0001-5984-197X} \and Walter Hugo Lopez Pinaya \inst{1}\orcidID{0000-0003-3739-1087}  \and Pedro Borges \inst{1}\orcidID{0000-0001-5357-1673} \and Mark S. Graham \inst{1}\orcidID{0000-0001-6435-5079} \and Tom Vercauteren \inst{1}\orcidID{0000-0003-1794-0456} \and M. Jorge Cardoso \inst{1}\orcidID{0000-0003-1284-2558}}
\authorrunning{Fernandez et al.}
%
\institute{King's College London, London WC2R 2LS, UK}
\maketitle              
\begin{abstract}
Generative modelling and synthetic data can be a surrogate for real medical imaging datasets, whose scarcity and difficulty to share can be a nuisance when delivering accurate deep learning models for healthcare applications. In recent years, there has been an increased interest in using these models for data augmentation and synthetic data sharing, using architectures such as generative adversarial networks (GANs) or diffusion models (DMs). Nonetheless, the application of synthetic data to tasks such as 3D magnetic resonance imaging (MRI) segmentation remains limited due to the lack of labels 
associated
with the generated images. Moreover, many of the proposed generative MRI models lack the ability to generate arbitrary modalities due to the absence of explicit contrast conditioning.
These limitations prevent the user from adjusting the contrast and content of the images and obtaining more generalisable data for training task-specific models. In this work, we propose brainSPADE3D, a 3D generative model for brain MRI and associated segmentations, where the user can condition on specific pathological phenotypes and contrasts. The proposed joint imaging-segmentation generative model is shown to generate high-fidelity synthetic images and associated segmentations, with the ability to combine pathologies. We demonstrate how the model can alleviate issues with segmentation model performance when unexpected pathologies are present in the data.

\end{abstract}
\section{Introduction}
In the past decade, it has been shown that deep learning (DL) has the potential to ease the work of clinicians in tasks such as imaging segmentation~\cite{Esteva2019}, an otherwise time-consuming task that requires expertise in the imaging modality and anatomy. Nonetheless, the performance and generalisability of DL algorithms is linked to how extensive and unbiased the training dataset is \cite{IanGoodfellowYoshuaBengio2015}. While large image datasets in computer vision are widely available \cite{imagenet}, this is not the case for medical imaging because images are harder to acquire and share, as they are subject to tight data regulations \cite{rieke}. In addition, most state-of-the-art segmentation algorithms are supervised and require labels as well as images, and because obtaining these 
requires substantial time and expertise,
they tend to focus on a specific region of interest, making dataset harmonisation harder. In brain MRI, where studies are tailored to a pathology and population of interest, obtaining a large, annotated, multi-modal and multi-pathological dataset is challenging. An option to overcome this is to resort to domain randomisation methods such as SynthSeg \cite{Billot2022}, but their performance in the presence of highly variable pathologies such as tumours has not been tackled. Alternatively, data augmentation via deep generative modelling, an unsupervised DL branch that learns the input data distribution, has been applied in recent years to enrich existing medical datasets, producing realistic, usable synthetic data with the potential to complement \cite{Barile2021} or even replace \cite{sashimi_brainspade} real datasets, using architectures such as generative adversarial networks (GANs) and the more recent diffusion models (DMs) \cite{Tudosiu,ldm_pinaya}.
One of the major roadblocks, though, when it comes to applying synthetic data to segmentation tasks is that of producing labelled data. Conditioning can give the user some control over the generated phenotypes, such as age \cite{ldm_pinaya}. However, to our knowledge, only a handful of works deliver segmentations to accompany the data. In the case of published models generating data based on real labels \cite{Qasim2020}, we must consider that labels are not usually shared, as they are sometimes considered protected health information, due to the risk of patient re-identification \cite{brainprint}, and due to the above-mentioned difficulty to produce. Therefore, it may be beneficial that the labels themselves are also algorithmically generated from a stochastic process. Few works in the literature provide synthetic segmentations \cite{Basaran2022,Foroozandeh2020,sashimi_brainspade}, especially enclosing healthy and multiple diseased regions. Among the latter, these models are limited to 2D, and they hardly allow the user to modulate their content (e.g., selecting the subject's pathology or age), limiting their applicability. 

\textbf{Contributions: } in this work, we propose a 3D generative model of the brain that provides multi-modal brain MR images and corresponding semantic maps generated by giving the user the power to condition on the pathological phenotype of the synthetic subject. We showcase the benefits of using these synthetic datasets on a downstream white matter hyperintensity (WMH) segmentation task when the test dataset contains also contains tumours. 

\section{Methods}

\subsection{Data}

For training, we used the SABREv3 dataset consisting of 630 T1, FLAIR and T2 images \cite{Jones2020}, a subset of 66 T1 and FLAIR volumes from ADNI2 \cite{Jack2008}, and 103 T1, FLAIR and T2 volumes from a set of sites from BRATS~\cite{Bakas2017,Bakas2018,Menze2015}. Due to the large computational costs associated with training generative models, it is not tenable to train them using full resolution, full size 3D images: we circumvented this issue by, on one hand, mapping images to a $2mm$ isotropic MNI space, resulting in volumes of dimensions $96\times128\times96$, and on the other hand, by operating with patches, taken from $1mm$ data of dimensions $146\times176\times112$.
Bronze-standard partial volume (PV) maps of the cerebrospinal fluid~(CSF), grey matter (GM), white matter (WM), deep grey matter~(DGM) and brainstem were obtained using GIF~\cite{cardoso2015geodesic}, masking out  tumours for BRATS. These healthy labels were overlaid with manual lesion labels provided with the datasets: WMH for the first two; and gadolinium-enhancing (GDE), non-enhancing (nGDE) tumour and edema for BRATS.

\subsection{Algorithm}
\begin{figure}[b!]
\centering
\includegraphics[width=0.95\textwidth]{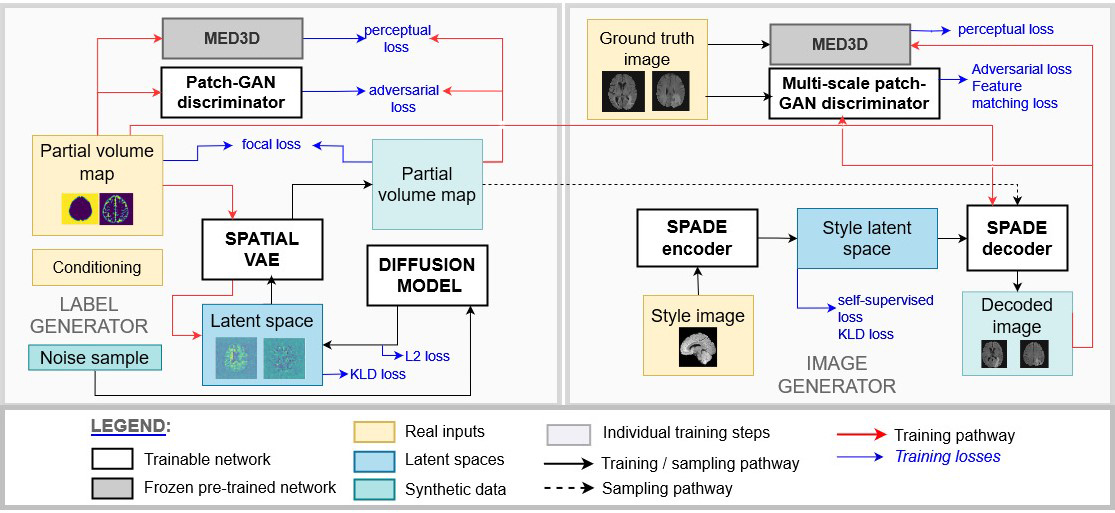}
\caption{Architecture of our two-stage model: the left block corresponds to the label generator, and the right block to the image generator. Training and inference pathways are differentiated with black, red and dashed arrows.} \label{fig:brainspade3d}
\end{figure}

Our pipeline consists of a conditional generator of semantic maps and an image generator, depicted in Fig. \ref{fig:brainspade3d}. It is based on the generative model proposed in \cite{sashimi_brainspade}: a label generator, consisting of a latent diffusion model (LDM), is trained on the healthy tissue and lesion segmentations. Independently, a SPADE-like \cite{spade} network is trained on the PV maps and the multi-modal images. For the $1mm^3$ data, patches of  $146\times176\times64$ had to be used for the image generator.

\textbf{Label generator}: The proposed label generator is based on a latent diffusion model (LDM), made up of a spatial variational auto-encoder (VAE) and a diffusion model (DM) operating in its latent space. The VAE of the $2mm^3$ resolution model had 3 downsamplings, and the $1mm^3$ had 4, resulting in latent spaces of shapes $32\times24\times32$ and $24\times24\times16$, respectively. The VAE is trained using focal loss ($\gamma=3$), Kullback-Leibler distance (KLD) loss to stabilise the latent space, Patch-GAN adversarial loss and a perceptual loss based on the features of MED3D \cite{Chen}, implemented using MONAI \cite{MONAI}. For the diffusion model, we predict the velocity using the \textit{v-parametrization} approach from \cite{Salimans} and optimise it via an $l_2$ loss. We used T=1000 timesteps. We used a PNDM \cite{Liu2022} scheduler to sample data, predicting only 150 timesteps. In addition, disease conditioning $dc$ was applied using a cross-attention mechanism. For each subject $j$ and disease type $l$, we have a label map $M_{jl}$, from which we produce a conditioning value $dc_{jl}$ reflecting the voxels labelled as $l$ in the map, normalised by the maximum number of $l$ voxels across the dataset, i.e.:
\begin{equation}
dc_{jl} = \frac{\sum_{n=1}^{N}M_{jl}}{max_j\sum_{n=1}^{N}M_{jl}},
\end{equation} 
where the sum is across all voxels in the map. We trained the VAE for 250 epochs and the DM for 400, on an NVIDIA A100 DGX node.

\textbf{Image generator}: We modified the SPADE model used in \cite{sashimi_brainspade} to extend the 2D generator and multi-scale discriminator to 3D. The encoder, which should only convert the contrast of the input image to a style vector, was kept as a 2D network, as it was found to work well while being parsimonious. To ensure that the most relevant brain regions informed the style, we used sagittal slices instead of axial ones as done in \cite{sashimi_brainspade}, selecting them randomly from the central 20 slices of each input volume. We kept the original losses from \cite{spade} to optimise the network. We replaced the network on which the perceptual loss is calculated with  MED3D \cite{Chen}, as its features are also in 3D and fine-tuned on medical images, which are more domain-pertinent. We ran a full ablation study on the losses introduced in \cite{sashimi_brainspade}; we dropped the modality-dataset discriminator loss, as it did not lead to major improvements but kept the slice consistency loss. To train the $1mm^3$ model, we used random patches of size 64 along the axial dimension. During inference, we used a sliding-window approach with a 5-slice overlap. We trained the networks for 350 epochs on an NVIDIA A100 GPU. Further details are provided in the supplementary materials. Code is available at \url{https://github.com/virginiafdez/brainSPADE3D_rel.git}. 

\subsection{Downstream segmentation task}
To compare the performance in segmentation tasks of our synthetic datasets, we performed several experiments using nnUNetv2 \cite{Isensee2021} as a strong baseline architecture, adjusting only the number of epochs until convergence. The partial volume maps were converted to categorical labels via an $\textit{argmax}$ operator. 

\section{Experiments}

\subsection{Quality of the generated images and labels pairs} \label{exp_quality}
\begin{figure}[h!]
\centering
\includegraphics[width=0.99\textwidth]{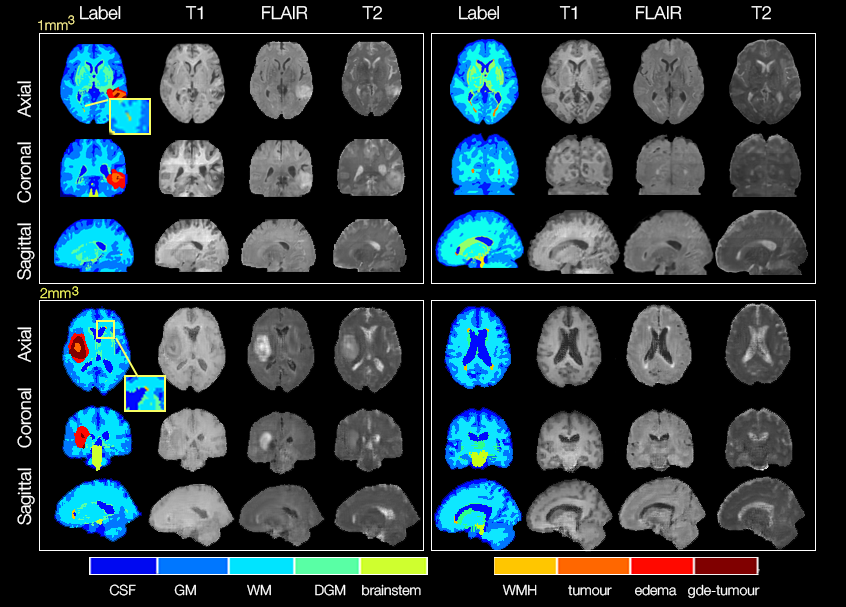}
\caption{Example synthetic $1mm^3$ and $2mm^{3}$ isotropic labels and images generated using tumour+WMH (left) and WMH (right) conditioning. The augmented frame in the top left images shows the small WMH lesions near the ventricles.} \label{fig:samples_grid}
\end{figure}

Without established baselines for paired 3D healthy and pathological labels and image pairs, we assess our synthetic data by comparing them to real data and showing how they can be applied to downstream segmentation tasks. 
Examples of generated labels and images are depicted in Fig.~\ref{fig:samples_grid}. In one case, we used a conditioning unseen by the model during training, \textit{WMH + all tumour layers}: both $1mm^3$ and $2mm^3$ label generators show the capability of handling  this unseen combination, resulting in both lesions being present in the resulting images. However, we observed a lower ability of extrapolating to unseen phenotypes in the $1mm^3$ model, with only about 37\% of the labels inferred using such conditioning resulting in the desired phenotype being met, as opposed to the $2mm^3$ model which manifested both lesion types in 100\% of the generated samples.

\noindent\textbf{Quality of the labels}: As the label generator is stochastic, we cannot compute paired similarity metrics. Instead, we compare the number $V_{i,j}$, for image $i$ and region $j$, of  CSF, GM, WM, DGM and brainstem voxels across subjects between a synthetic dataset of 500 volumes and a subset of the training dataset of the same size, excluding tumour images, but allowing for low WMH values as these don't disrupt the anatomy of the brain. The number of voxels $V_{i,j}$ is calculated as: $V_{i,j}= \sum_{n=1}^{N} (i_{n==j})  $
, where N is the number of pixels in the image. Table \ref{proportion_tissue} reports mean values and standard deviations, demonstrating that our model generates labels with mean volumes similar to real data. By comparing the labels, we saw that the considerable discrepancy between $V_{i, CSF}$ at $1mm^3$ is due to a loss of details in the subarachnoid CSF, which is very thin, likely due to the high number of VAE downsamplings at $1mm^3$ (visual comparison is available in supplementary Fig. 1).  However, we observe a lower standard deviation in the tissue volumes, indicating that the label generator does not capture the natural variability of brain tissues.

\begin{table}[t!]
\centering
\caption{Mean number of voxels and standard deviation of brain regions across real and synthetic datasets. Every value has been multiplied by $10^{-4}$.} 
\begin{tabular}{|c|ccccc|}
\hline
\textbf{Dataset} & \textbf{CSF} & \textbf{GM} & \textbf{WM} & \textbf{DGM} & \textbf{Brainstem}  \\
\hline
Real ($1mm^3$) & $20.342_{2.830}$ & $37.634_{1.747}$ & $44.872_{2.270}$ & $4.301_{0.581}$ & $1.238_{0.269}$ \\
Synthetic ($1mm^3$) & $14.141_{0.964}$ & $43.583_{0.735}$ & $42.029_{1.728}$ & $6.713_{0.592}$ & $1.066_{0.102}$ \\
\hline
\hline
Real ($2mm^3$) & $4.790_{0.573}$ & $9.879_{0.585}$ & $7.436_{0.512}$ & $0.453_{0.080}$ & $0.363_{0.036}$ \\
Synthetic ($2mm^3$) & $4.641_{0.163}$ & $8.564_{0.297}$ & $7.008_{0.198}$ & $0.587_{0.978}$ & $0.343_{0.013}$ \\
\hline
\end{tabular}
\label{proportion_tissue}
\end{table}

\noindent\textbf{Quality of the generated images}: To assess the performance of our image generator, we use a hold-out test set of PV maps and corresponding T1, FLAIR and T2 images, to compute the structural similarity index (SSIM) between the ground truth images and the image generated when the real PV map and a slice from the ground truth were used as inputs to the models. The mean SSIMs obtained are summarised in table \ref{image_mets}, showing that the model performs similarly across different contrasts. Discrepancies between real and synthetic images can be explained by the stochasticity of the style encoder.

\begin{table}[b!]
\centering
\caption{SSIM values obtained between real and synthetic images for T1, FLAIR and T2 contrasts for both models, generated using real PV maps.} 
\begin{tabular}{|c|c|c||c|c|c|}
\hline
\multicolumn{3}{|c||}{$1mm^3$} & \multicolumn{3}{|c|}{$2mm^3$} \\
\hline
T1 & FLAIR & T2 & T1 & FLAIR & T2 \\
\hline
$0.842_{0.083}$ & $0.798_{0.082}$ & $0.794_{0.075}$ & $0.922_{0.010}$ & $0.910_{0.030}$ & $0.909_{0.025}$ \\
\hline
\end{tabular}
\label{image_mets}
\end{table}

\label{neurotypical}
\begin{table}[t!]
\caption{Mean Dice score and standard deviation obtained for the models trained on real and synthetic data. Asterisks denote significantly better performance.}
\centering
\begin{tabular}{|c|c||c|c|c|c|c|}
\hline
\textbf{Resolution} & \textbf{Model} & \textbf{CSF} & \textbf{GM} & \textbf{WM} & \textbf{DGM} & \textbf{Brainstem} \\
\hline
$1mm^3$ & $M_{healthy}$ & $0.957_{0.005}$* & $0.959_{0.003}$* & $0.971_{0.003}$* & $0.875_{0.015}$*& $0.958_{0.021}$* \\
$1mm^3$ & $M'_{healthy}$ & $0.884_{0.014}$ & $0.912_{0.009}$ & $0.936_{0.005}$ & $0.684_{0.034}$ & $0.874_{0.036}$ \\
\hline
$2mm^3$ & $M_{healthy}$ & $0.947_{0.057}$* & $0.958_{0.046}$* & $0.968_{0.039}$* & $0.887_{0.065}$*& $0.962_{0.024}$* \\
$2mm^3$ & $M'_{healthy}$ & $0.869_{0.057}$ & $0.895_{0.052}$ & $0.931_{0.047}$ & $0.703_{0.100}$ & $0.905_{0.025}$ \\
\hline
\end{tabular}
\label{neurotypical_tissue}
\end{table}



\noindent\textbf{Synthetic data for healthy region segmentation}:
We assess the performance of our generated pairs on a CSF, GM, WM, DGM and brainstem segmentation task. We train an nnU-Net model, $M_{healthy}$ on the T1 volumes of the real subset of 500 subjects mentioned earlier, and $M'_{healthy}$ on the 500 synthetic T1 and label pairs, then test both models on a hold-out test set of 30 subjects from the SABRE.  The Dice scores on all regions are reported in table~\ref{neurotypical_tissue}. Although $M_{healthy}$ performs better in all regions, the $M'_{healthy}$ trained on purely synthetic data demonstrated a competitive performance for all regions except the DGM. DGM is a complex anatomical region comprising several small structures with intensities ranging between those typical for GM and WM. Thus, the PV map value for a voxel in this region will split its probability between DGM, WM and GM rather than favouring just one class, which is problematic for nnU-Net, as it requires categorical inputs to train the model, resulting in noisy ground truth labels that cause a larger distribution shift for the region. Examples of these noisy training and test labels are showcased in supplementary Fig. 2. 


\subsection{WMH segmentation in the presence of tumour lesions}

\textbf{Aim}: The main aim of this work is to show how synthetic data can increase the performance of segmentation models when training datasets are biased towards a specific phenotype. We focus on WMH segmentation. Our target dataset is a subset of 30 unseen volumes from BRATS (a different set of sites from those seen by the generative model) containing WMH lesions and tumours. We hypothesise that a WMH segmentation model trained on images that do not contain tumours will label these as WMH, as tumours and WMH share some intensity similarities in the FLAIR contrast typically used to segment WMH \cite{Sudre2015}. With the proposed generative model, we can generate synthetic data containing subjects with \textit{both} tumours and WMH, which should make the training model robust to cases where both diseases are present, therefore reducing false positives.

We ran this experiment with $2mm^3$ isotropic data, as the phenotype conditioning worked better (see \ref{exp_quality}) in this model. From a stack of 500 real FLAIR volumes from the SABRE dataset, and a stack of 500 FLAIR synthetic volumes generated from synthetic labels conditioned on both tumours and WMH, we train several models $M_{R_{PR}S_{PS}}$, varying the \% proportions $PR$ and $PS$ of real and synthetic data respectively. In addition, even if the premise of this work is that users do not have access to real data containing tumours, we train model $M_{R_{WMH}R_{tum}}$ on the real FLAIR volumes from the SABRE dataset, \textit{and} the BRATS tumour volumes used to train the synthetic model, leaving the training labels empty, as no prior WMH segmentations are available for BRATS. We calculate the Dice score on WMH on a hold-out test set of 30 subjects from the SABRE dataset. As we do not have WMH ground truth labels for our test set from BRATS, but we have tumour labels, we compute the proportion of tumour pixels incorrectly labelled as WMH and note this metric $FP_{tum}$. 

\begin{table}[b!]
\caption{Mean Dice score, precision and recall obtained on the SABRE test dataset by all the WMH segmentation models we trained, and $FP_{tum}$ ratio on the BRATS holdout dataset. Asterisks indicate statistical significance.}
\centering
\begin{tabular}{|c||c|c|c|c|}
\hline
\textbf{Model} & \textbf{Dice (PD) $\uparrow$} &  \small\textbf{precision (PD) $\uparrow$} & \textbf{recall (PD) $\uparrow$} & \textbf{$FP_{tum}$ \small{(BRATS)} $\downarrow$} \\
\hline
\textbf{$M_{R_{100}S_{0}}$} & $0.728_{0.281}$* & $0.761_{0.236}$ & $0.751_{0.132}$*& $0.325_{0.232}$ \\
\hline
\textbf{$M_{R_{75}S_{25}}$} & $0.713_{0.208}$ & $0.745_{0.231}$ & $0.743_{0.137}$&$0.902_{0.187}$  \\
\hline
\textbf{$M_{R_{50}S_{50}}$} & $0.716_{0.211}$* & $0.742_{0.227}$ & $0.754_{0.132}$* & $0.108_{0.209}$\\
\hline
\textbf{$M_{R_{25}S_{75}}$} & $0.722_{0.210}$ & $0.755_{0.225}$ & $0.722_{0.138}$ &
$0.079_{0.171}$\\
\hline
\textbf{$M_{R_{5}S_{95}}$} & $0.642_{0.210}$& $0.716_{0.243}$& $0.628_ {0.146}$&$0.023_{0.078}$\\
\hline
\textbf{$M_{R_{0}S_{100}}$} & $0.362_{0.147}$ & $0.726_{0.294}$ & $0.263_{0.104}$&$0.001_{0.002}$*\\
\hline
\textbf{$M_{R_{WMH}R_{tum}}$} & $0.710_{0.233}$ & $0.742_{0.250}$ & $0.741_{0.131}$ &$0.026_{0.140}$*\\
\hline
\end{tabular}
\label{results_tumour}
\end{table}

\begin{figure}[t!]
\centering
\includegraphics[width=1\textwidth]{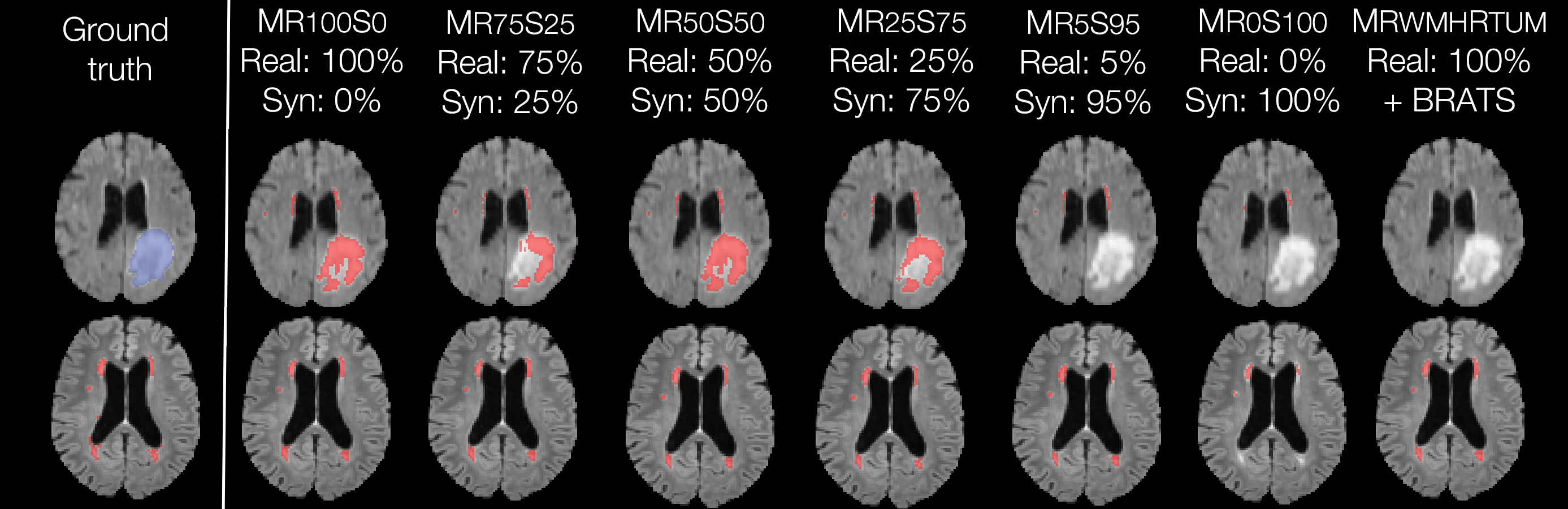}
\caption{Sample WMH predictions on the BRATS dataset (top) and the SABRE test set (bottom) for all our models, in red. The leftmost column shows the tumour mask for the BRATS dataset (in blue) and the ground truth WMH for SABRE.  
} \label{fig:grid_wmh}
\end{figure}

\textbf{Results}:  Results are reported in table~\ref{results_tumour}. Example segmentations and predicted WMH masks on both test sets are depicted in Fig.~\ref{fig:grid_wmh}. $M_{R_{100}S_{0}}$ achieved the top Dice on WMH for its in-domain test set, but it has one of the worse $FP_{tum}$ scores on the BRATS set. While $M_{R_{0}S_{100}}$ achieved a low Dice on the SABRE test set, it had a $FP_{tum}$ score that is significantly lower than that of $M_{tum-real}$. All the models trained on a combination of real and synthetic data achieve a competitive $FP_{tum}$ without compromising the WMH Dice. While all models have a comparable precision, $M_{R_{0}S_{100}}$ has low recall; caused by an underestimation of WMH, as seen in Fig.~\ref{fig:grid_wmh}. Interestingly, besides $M_{R_{0}S_{100}}$ and $M_{R_{5}S_{95}}$, the edematous area of the tumours still gets partially segmented as WMH. $M_{R_{WMH}R_{tum}}$ achieves very good Dice, precision, recall and $FP_{tum}$ metrics; but, while examining the WMH segmentations on the BRATS dataset, all the segmentations were empty, as shown in Fig. \ref{fig:grid_wmh}, which indicates that, because the WMH training labels for BRATS were empty, the model has mapped the appearance and/or phenotype of the BRATS dataset to an absence of WMH.

\section{Discussion \& Conclusion}

This work presents a label and multi-contrast brain MRI 3D image generator that can supplement real datasets in segmentation tasks for healthy tissues and pathologies. The synthetic data provided by our model can boost the precision and robustness of WMH segmentation models when tumours are present in the target dataset, showing the potential for having content and style-disentangled generative models that can combine the phenotypes seen in their training datasets. While disentanglement is covered in \cite{sashimi_brainspade}, 3D can help produce data usable in scenarios where the context of neighbouring slices is meaningful, such as segmenting small lesions. In addition, disease conditioning, which was not implemented in \cite{sashimi_brainspade}, can be challenging in 2D, as some diseases depend on the axial location, such as WMH. 
Our current set-up has, however, some limitations. First, there is a caveat in using $2mm^3$ isotropic data or patching at $1mm^3$. Even so, diffusion models for high resolution 3D images have to operate in a latent space that causes loss of semantic variability (See \ref{exp_quality}) and small details, affecting the downstream segmentation task, partly because a gap appears in the image synthesis process between synthetic and real labels. Further work should attempt to harmonise the semantic synthetic and real domains. Although one of the causes of this limitation is capacity, the latest advances in diffusion models show that higher performance and resolution can be achieved \cite{Hoogeboom}, potentially leading to better labels and, effectively, less domain shift between real and synthetic domains. Secondly, conditioning on variables such as age or ventricle size could also translate into more variability across the generated volumes \cite{ldm_pinaya}, overcoming the limitation in tissue variability observed in table \ref{proportion_tissue}. The method can be scaled to more pathologies and tasks, as model sharing allows for fine-tuning on more pathological labels, thus making segmentation models more generalisable to the diverse phenotypes of real brain MR data. 


%
%
%
%
\bibliographystyle{splncs04}
\bibliography{bibliography.bib}
\end{document}


\appendix

\section{Training hyperparameters and augmentations}
\begin{table}[ht!]
\centering
\caption{Label and image generator hyperparameters}
\newcolumntype{Y}{>{\centering\arraybackslash}X}
\resizebox{0.95\textwidth}{!}{
\begin{tabularx}{\linewidth}{|Y|Y||Y|Y|} 
\hline
\multicolumn{4}{|c|}{\textbf{Variational autoencoder}} \\
\hline
\multicolumn{2}{|c||}{\textbf{Loss weights}} & \multicolumn{2}{|c|}{Other parameters} \\
\hline
Patch GAN loss & 0.1 & epochs & 250 \\
\hline
KLD loss & $10^{-8}$ & train time & 4h38 min \\ 
\hline
reconstruction loss &  1.0 & optimiser & Adam  \\ 
\hline
perceptual loss & 10 & learning rate & vae: $2\cdot10^{-4}$ disc: $10^{-4}$   \\ 
\hline
hardware &  NVIDIA DGX A100 & batch size & 8 \\
\hline
\multicolumn{4}{|c|}{\textbf{Latent diffusion model}} \\
\hline
epochs & 400 & training time & 12h \\
\hline
warm-up learning rate & $10^{-8}$ &  batch size & 8 \\ 
\hline
base learning rate & $2.5\cdot10^{-5}$ & loss  & $l_1$  \\ 
\hline
perceptual loss & 10 & hardware & NVIDIA DGX A100    \\ 
\hline
\hline
\multicolumn{4}{|c|}{\textbf{Image generator}} \\
\hline
Epochs & 350 & training time & 2 weeks \\
\hline
hardware & single A100 GPU & optimiser & Adam \\ 
\hline
learning rate &  $2\cdot10^{-4}$ & number of discriminator & 3  \\ 
\hline 
\multicolumn{4}{|c|}{Loss weights} \\
\hline
\multicolumn{2}{|c||}{\textbf{Generator}} & \multicolumn{2}{|c|}{\textbf{Discriminator}} \\
\hline
KLD & $10^{-5}$ & feature loss & 0.25 \\
\hline
perceptual loss & 1.5 &   \multicolumn{2}{|c|}{\textbf{gen. and disc. training thresholds}} \\ 
\hline
slice consistency & 0.5 &  lower (D only)  & 0.65  \\ 
\hline
perceptual loss & 10 &  lower (G only) &  0.75 \\ 
\hline
\end{tabularx}}
\end{table}


Augmentations were implemented using MONAI (https://monai.io/). 
\begin{center}
\begin{table}[ht!]
\centering
\caption{Transformations applied to the different modules. The intensity transforms, marked with (*) were only applied to images, not labels.}
\resizebox{0.99 \textwidth}{!}{
\begin{tabular}{|p{2.5cm}|p{5cm}|p{5cm}|p{4.5cm}|} 
\hline
Augmentation & VAE (ranges) & DM (ranges) & Image generator (ranges) \\ [0.5ex] 
\hline
Random affine & rotation: [-0.05, 0.05], shear: [0.001, 0.05], scale:[0, 0.05], probability: 0.15 & rotation: [-0.1, 0.1], shear: [0.001, 0.15], scale:[0, 0.3], probability: 0.15 & rotation: [-0.05, 0.05], shear: [0.001, 0.05], scale:[0, 0.05], probability: 0.33 \\ 
\hline
Random bias field (*) & - & - & intensity: (0, 0.005), probability: 0.33 \\
\hline
Random Gaussian noise (*) & - & - & mean: 0.0, $\sigma$ range: [0.005, 0.015], probability: 0.33 \\
\hline
Random contrast adjust (*) & -  & - & $\gamma$ range: [0.9, 1.15], probability: 0.33 \\
\hline
\end{tabular}}
\end{table}
\end{center}

\section{Additional result figures}

\begin{figure}[h!]
\centering
\includegraphics[width=0.70\textwidth]{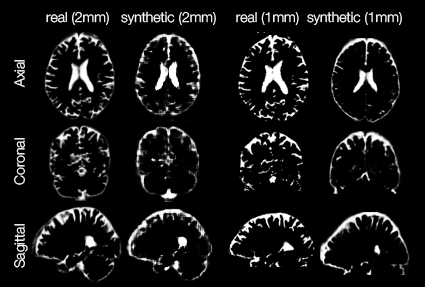}
\caption{Example CSF real and synthetic label channels for both the $1mm^3$ and $2mm^3$ models, showing the loss of detail in the subarachnoid space CSF on the $1mm^3$ model. All images are unpaired.} \label{fig:grid_wmh}
\end{figure}
\begin{figure}[b!]
\centering
\captionsetup{width=.99\textwidth}
\includegraphics[width=0.85\textwidth]{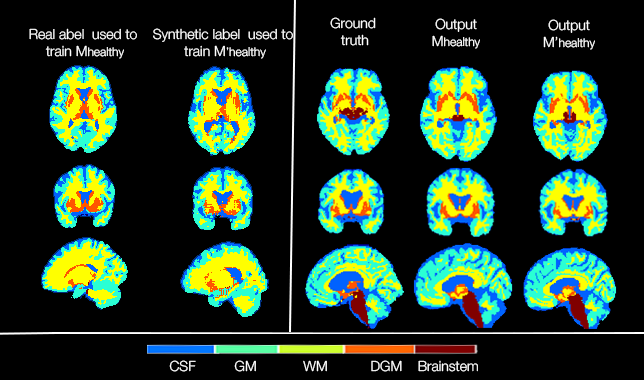}
\caption{From left to right: example real label from the SABRE dataset used to train model $M_{healthy}$ (see section 3.1), synthetic healthy label generated by our model used to train $M'_{healthy}$, ground truth sample from the test set of the SABRE dataset and corresponding $M_{healthy}$ and $M'_{healthy}$ outputs. Examples shown are for the $2mm^3$ model.} \label{fig:grid_wmh}
\end{figure}

%
%
%